\documentclass[intlimits,twoside,a4paper]{article}

\usepackage{amsmath,amssymb}
\usepackage{graphicx}

\usepackage[T2A]{fontenc}
\usepackage[cp1251]{inputenc}

\usepackage{cmpj2}
\issue{2016}{19}{4}{43701}
\doinumber{10.5488/CMP.19.43701}
\title[Properties and temperature evolution of the spectrum of localized quasi-particles]%
{Properties and temperature evolution of the spectrum of localized quasi-particles interacting with polarization phonons in two models}
\author[M.V.~Tkach, Ju.O.~Seti, O.M.~Voitsekhivska, O.Yu.~Pytiuk]{M.V.~Tkach\thanks{E-mail: ktf@chnu.edu.ua}\,,
Ju.O.~Seti, O.M.~Voitsekhivska, O.Yu.~Pytiuk}
\address{Chernivtsi National University, 2 Kotsyubinsky St.,
58012 Chernivtsi, Ukraine}
\authorcopyright{M.V.~Tkach, Ju.O.~Seti, O.M.~Voitsekhivska, O.Yu.~Pytiuk, 2016}
\date{Received April 26, 2016, in final form June 2, 2016}

\begin{document}

\maketitle

\begin{abstract}
Using the Feynman-Pines diagram technique, the energy spectrum of localized quasi-particles interacting with polarization phonons is calculated and analyzed in the wide range of energies at the finite temperature of the system. It is established that the general model of  the system, besides the bound states known from the simplified  model with an additional condition for the operator of quasi-particles number, contains the new bound states even for the systems with weak coupling. The contribution of multi-phonon processes into the formation of renormalized spectrum of the system is analyzed.  The reasons of the appearance, behaviour and disappearance of separate pairs of bound states depending on the coupling constant and temperature are revealed.

\pacs 71.38. - k, 63.20.kd, 63.20.dk, 72.10.Di
\end{abstract}

\section{Introduction}

The theory of the renormalized energy spectrum of a system of quasi-particles interacting with pho\-nons in a wide range of energies (containing bound states) and coupling constants has been attracting a permanent attention \cite{Mis05, Fil12}. The increased interest to the theory of quasi-particles interacting with the quantized fields (photons, phonons) is caused by the practical and fundamental prospects of their utilization.

As an example of practical use we should mention the important role of phonon- and photon assisted tunneling of electrons through the multi-layer resonant tunneling nanostructures (RTS), where, due to the interaction with the fields, the new states between which the quantum transitions occur are observed causing the appearance of new transmitting canals of the system \cite{Tka12, Tka13}. These phenomena essentially effect the functioning of nanodevices (quantum cascade lasers and detectors), the main operating elements of which are the multi-layer  RTS \cite{Wen08, Gio09, Sim11}.

Studying the high-temperature superconductivity, mezo-physical phenomena, low-dimensional nano\-systems one should solve the problems where the perturbation or variational methods cannot be used because the multi-phonon or multi-photon processes of quasi-particles scattering play an essential or even the main role.  Therefore, one should propose the new mathematical approaches or use the methods of quantum field theory in order to overcome the known problems of partial summing of the infinite ranges of Feynman diagrams. For example, within the new diagrammatic Monte Carlo method \cite{Pro98, Prok98, Mis00}, the important problems of high-temperature superconductivity \cite{Mis09} have been solved, and the excited states of structure and structureless Fr\"{o}hlich polaron \cite{Mis05, Fil12} have been studied exactly and consistently. Recently in \cite{Ber06, Ber10, Mar10, Ebr12}, a new powerful approach for the investigation of different quasi-particles interacting with phonons in homogeneous and inhomogeneous systems has been developed. It is based on the non-perturbative method of momentum average approximation. Using it, in the paper \cite{Ebr12}, all types of disorder for various types of electron-phonon interaction were studied in detail. Here, it was shown that the obtained results well correlate with that obtained within the diagrammatic Monte Carlo method. Moreover, in \cite{Tka15}, the modified method of Feynman-Pines diagram technique has been used for the development of a physically correct theory of renormalized spectrum of the ground and first excited state of Fr\"{o}hlich polaron in the regime of weak coupling at $T=0$ K taking into account the multi-phonon processes.

A complicated mathematical approach of quantum field theory does not always give an opportunity to study the spectra of quasi-particles renormalized due to the interaction with phonons in the framework of the general realistic models without the use of uncontrolled approximations. For several systems, this problem is studied using some additional mathematical conditions, which have a proper physical interpretation simplifying the starting model after which the problem is solved exactly.

The above mentioned situation is well known in the theory \cite{Dav76, Tka03} of  quasi-particles with narrow energy bands (or localized ones) interacting with polarization phonons, where the renormalized spectrum of the system is obtained using an additional condition for the operator of a complete number of quasi-particles, which is equal to the square of this operator. Physically, this condition is explained the one at which the quasi-particle either exists in a certain state or not, without considering the interaction with phonons. In this simplified model, the unitary transformation with further analytical calculation of Fourier image of quasi-particles Green's function allowed us to obtain the exact energy spectrum \cite{Dav76} of all bound states of the system. It was shown that the renormalized spectrum, besides the ground level shifted into low-frequency range, contains a set of equidistant levels (the distance is of one phonon energy order) both in low- and high-frequency range independently of temperature.

In this paper, we obtain a renormalized spectrum of localized quasi-particles interacting with polarization phonons using the general model (without the above mentioned additional condition) for systems with weak coupling at a finite temperature. This problem is solved within the modified method of Feynman-Pines diagram technique. It is shown that compared to the simplified model, the general model provides a richer energy spectrum containing new bound states with complicated behaviour depending both on the temperature and coupling constant.

\section{Hamiltonian of the system. Theory of renormalized energy spectrum in two models}

We consider the localized dispersionless quasi-particles (excitons, impurities and so on) interacting with dispersionless polarization phonons described by Fr\"{o}hlich Hamiltonian

\begin{equation} \label{GrindEQ__1_}
H=E\sum _{\vec{k}}{\widehat{a}}_{\vec{k}}^{+} {\widehat{a}}_{\vec{k}}+{\Omega}\sum _{\vec{q}}\left(\widehat{b}_{\vec{q}}^{+} {\widehat{b}}_{\vec{q}} +\frac{1}{2}\right)+\sum _{\vec{k}\vec{q}}\varphi(q)\widehat{a}_{\vec{k}+\vec{q}}^{+}{\widehat{a}}_{\vec{k}}\left({\widehat{b}}_{\vec{q}} +{\widehat{b}}_{-\vec{q}}^{+}\right),
\end{equation}
where $E$ and  $\Omega$ are the energies of quasi-particles and phonons, respectively. In the range of their typical magnitudes for the bulk crystals and nano-systems ($E\approx 100\div 1000$~meV, $\Omega \approx 20\div 100$~meV), these values can be arbitrary. The analytical form of the binding function $\varphi(q)$ is not essential because the energy of quasi-particle-phonons interaction, as would be seen further, is consistently characterized by a coupling constant $\alpha $, independent of $\vec{q}$, which can be also arbitrary in natural range since the problem becomes zero-dimensional. The quasi-particle operators of the second quantization ($\widehat{a}_{\vec{k}},\widehat{a}_{\vec{k}}^{+}$) and ($\widehat{b}_{\vec{q}}, \widehat{b}_{\vec{q}}^{+} $) satisfy the Bose commutative relationships.

We study the problem of a renormalized spectrum of the system of quasi-particles interacting with phonons at arbitrary temperature ($T$) using the Hamiltonian  \eqref{GrindEQ__1_} within two models.

a) The model of the system where the condition
\begin{equation} \label{GrindEQ__2_}
\widehat{n}^{2} =\widehat{n}=\sum _{\vec{k}}\widehat{a}_{\vec{k}}^{+}  \widehat{a}_{\vec{k}}
\end{equation}
is fulfilled \cite{Dav76}. It means that the eigenvalue of these operators can be either 0 or 1 and is assumed in  \cite{Dav76} to be the condition of existence (1) or absence (0) of ``pure'' quasi-particle state (without interaction with phonons).

b) The general model of the system where the number of quasi-particle states is not confined by any conditions.

It is well known,  \cite{Dav76, Tka03, Abr12}, that the renormalized energy spectrum of the system is obtained from the Fourier image of the quasi-particle Green's function  $G(\vec{k},\omega')$, in general case, through the Dyson equation
\begin{equation} \label{GrindEQ__3_}
G(\vec{k},\omega')=\left[\omega'-E-M(\vec{k},\omega')\right]^{-1},\qquad(\hbar =1,\,\,\omega'=\omega +\ri\eta),
\end{equation}
related with the mass operator (MO) $M(\vec{k},\omega')$, which is calculated according to the rules of Feynman-Pines diagram technique \cite{Dav76, Tka03} when the concentration of quasi-particles is small.

In the model (a) the Hamiltonian \eqref{GrindEQ__1_} is diagonalized  exactly, using the known,  \cite{Dav76}, unitary transformation, and the poles of $G(\vec{k}, \omega')$ function are obtained, in its turn, defining the exact renormalized energy spectrum of the system
\begin{equation} \label{GrindEQ__4_}
E_{n}=E-\sum_{\vec{q}}\frac{|\varphi(q)|^{2}}{\Omega}+n\Omega,\qquad(n=0,\,\pm1, \,\pm2,\ldots).
\end{equation}
Formula \eqref{GrindEQ__4_} proves that the spectrum of the system is stationary (without a decay), equidistant and independent of temperature. It contains a renormalized by interaction ground state with the energy $E_{0}=E-\sum _{\vec{q}}|\varphi(q)|^{2}/\Omega$ shifted into the long-wave region with respect to $E$ and bound high-energy ($E_{n\geqslant1}$) and low-energy ($E_{n\leqslant-1}$) states of the system.

The exact unitary transformation, that makes the Hamiltonian \eqref{GrindEQ__1_} diagonal, cannot be performed for the model (b), where the condition \eqref{GrindEQ__2_} is not fulfilled. Thus, one should use the Feynman-Pines diagram technique \cite{Dav76, Tka03} in order to calculate $G(\vec{k}, \omega')$. All energetic multipliers in MO are independent of  quasi-momentum because the energies  ($E$, ${\Omega }$) are dispersionless, as far as they are carried out from all sums over~$\vec{q}$. Consequently, in all MO terms, one can see the same value $\sum_{\vec{q}}|\varphi(q)|^{2}/\Omega$, essentially simplifying the diagram technique \cite{Tka03}. Within the convenient dimensionless variables and values
\begin{equation} \label{GrindEQ__5_}
\xi=\frac{\omega-E}{\Omega}\,, \qquad\alpha=\sum _{\vec{q}}\frac{|\varphi(q)|^{2} }{\Omega^{2} }\,,\qquad m=\frac{M}{\Omega}\,,\qquad g=\Omega G,
\end{equation}
the Dyson equation \eqref{GrindEQ__3_} is of the form
\begin{equation} \label{GrindEQ__6_}
g(\xi)=\left[\xi-{\Large\textsl{m}(\xi)}\right]^{-1}.
\end{equation}
Here, since the problem has become ``zero-dimensional'', MO ${\Large\textsl{m}}(\xi)$ is defined by the infinite sum of ``dumb'' diagrams (without arrows at the dashed phonon lines and without numeric indices at the solid quasi-particles ones)
\begin{equation} \label{GrindEQ__7_}
\hfil\includegraphics[width=5.2in]{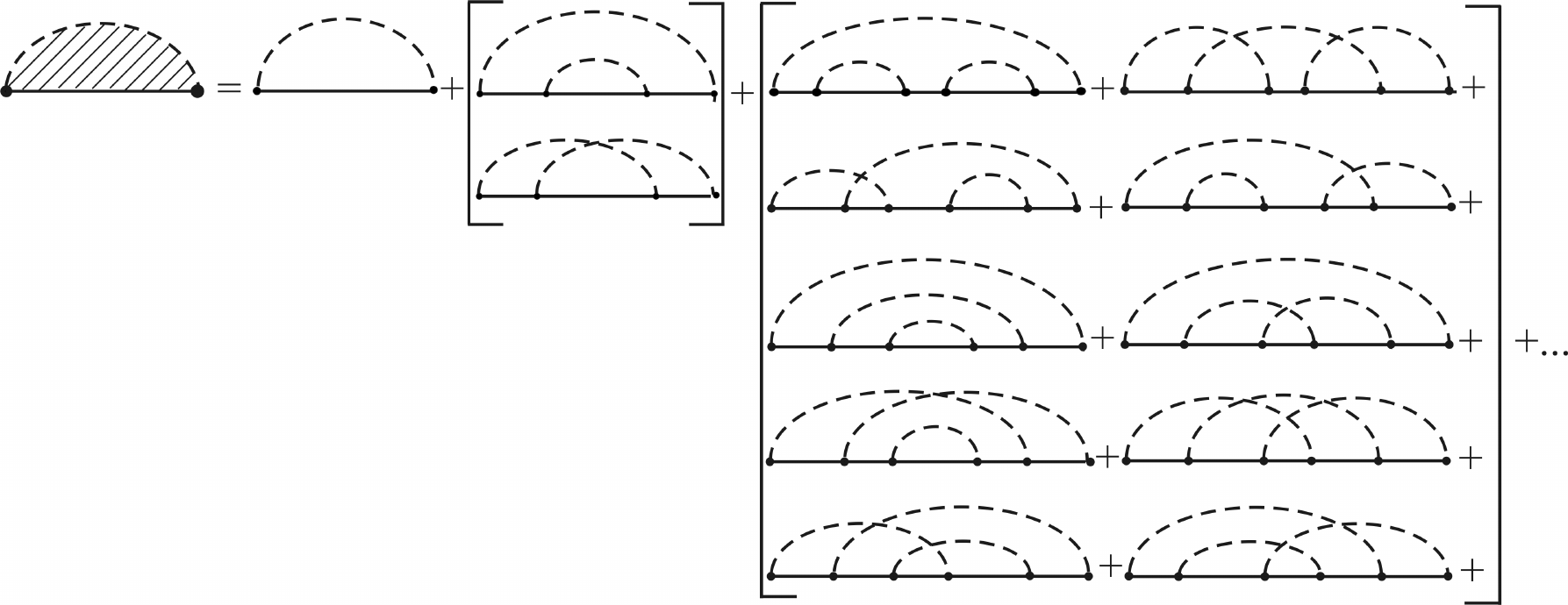}
\end{equation}
with the known rules of their construction, from the diagram technique  \cite{Tka03}.

In order to write a consistent analytic expression with respect to MO \eqref{GrindEQ__7_}, for each diagram of $n$-th order over the number of dashed lines, one should put in correspondence the sum of $2^{n}$ equal ``dumb'' diagrams with all possible directions of the arrows. Under the solid lines of these diagrams one should put the numbers, which are the sums of the numbers of dashed lines in the right-hand side (with sign ``$-$'') and in the left-hand side (with sign ``$+$''), placed above the respective solid line. For example, the system of four diagrams with indices
\begin{equation} \label{GrindEQ__8_}
\hfil\includegraphics[width=5.2in]{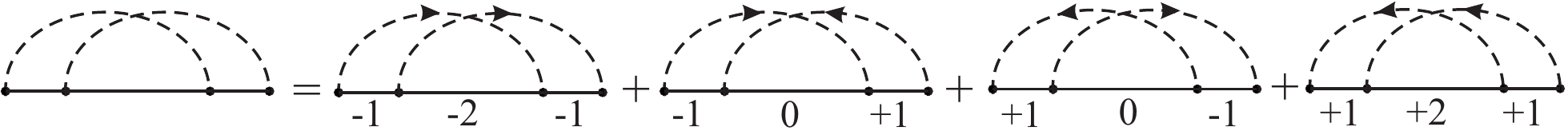}
\end{equation}
corresponds to the ``dumb'' diagram of the second order. The rules of conformity between diagrams and analytical expressions are simple:
\begin{equation} \label{GrindEQ__9_}
\raisebox{0pt}{\hfil\includegraphics[width=5.2in]{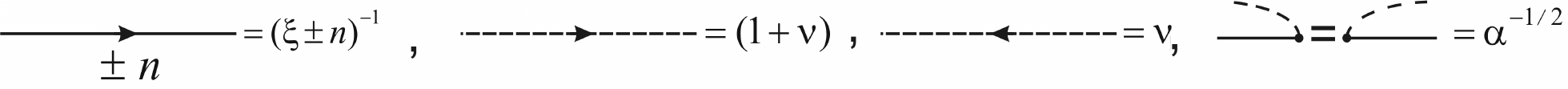}}\raisebox{8pt}{$\,\,,$}
\end{equation}
where $\nu =(\re^{\Omega /kT} -1)^{-1} $ is an average phonon occupation number.

Thus, the analytical expression for the arbitrary diagram with indices is written as a product of contributions of all its lines and tops. For example, the analytical expression
\begin{equation} \label{GrindEQ__10_}
\hfil\includegraphics[width=2.8in]{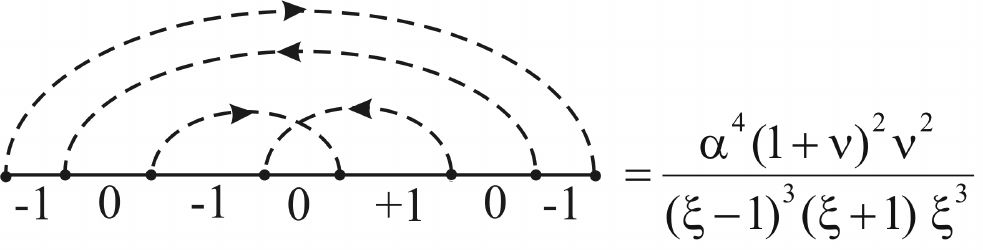}
\end{equation}
corresponds to the index diagram of the third order. The analysis of the structure of the complete MO proves that it can be written in an analytical form
\begin{equation} \label{GrindEQ__11_}
{\Large\textsl{m}}(\xi )={\Large\textsl{m}}_{1} (\xi )+\sum _{l=1}^{\infty }\left[{\Large\textsl{m}}_{2l} (\xi )+{\Large\textsl{m}}_{2l+1} (\xi )\right]
\end{equation}
with separated terms which describe the processes of quasi-particle scattering at phonons in the unmixed [${\Large\textsl{m}}_{1} (\xi )={\Large\textsl{m}}_{1}^{-} (\xi )+{\Large\textsl{m}}_{1}^{+} (\xi )$] and all possible mixed (the terms in $\sum_{\overrightarrow{q}}$) sequences. Herein, the term ${\Large\textsl{m}}_{1}^{-} (\xi )$ describes the interaction between quasi-particle and phonons in all unmixed processes, accompanied at first by the creation and then by the annihilation of phonons, while the term  ${\Large\textsl{m}}_{1}^{+} (\xi )$ --- vice versa. The sum of the rest of the terms in MO \eqref{GrindEQ__11_} describes the mixed processes of the phonon creation and annihilation in all possible sequences.

Therefore, MO ${\Large\textsl{m}}_{1}^{-} (\xi )$, according to the Feynman-Pines diagram technique, is an infinite sum of diagrams in all orders over the powers of $\alpha$ with the arrows directed only to the right (~\includegraphics[width=0.45in]{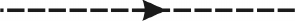}~$=1+\nu$). Therein, the diagrams with the crossing dashed lines give the same contribution as their equivalent diagrams without these crossing lines. Thus, ${\Large\textsl{m}}_{1}^{-} (\xi )$ has a rather simple diagram representation
\begin{equation} \label{GrindEQ__12_}
\raisebox{0pt}{\hfil\includegraphics[width=5.2in]{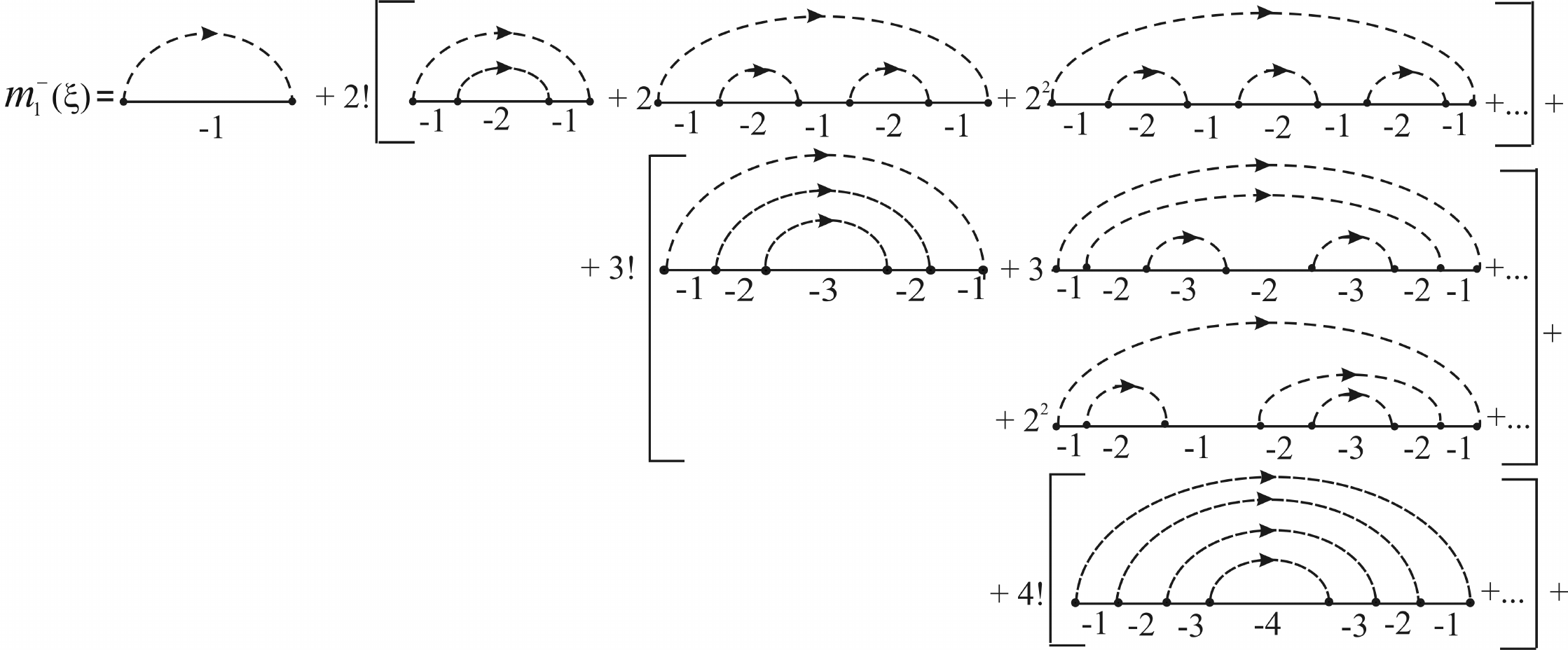}}\raisebox{10pt}{$\,\,\ldots\,.$}
\end{equation}

According to the rules of Feynman-Pines diagram technique, it brings us to the exact representation of this MO in two equal but analytically different forms
\begin{equation} \label{GrindEQ__13_}
{\rm {\Large\textsl{m}}}_{1}^{-} (\xi )=\frac{\alpha (1+\nu )}{\xi-1-\displaystyle\frac{2 \alpha (1+\nu )}{\xi -2-\ldots-\displaystyle\frac{n \alpha  (1+\nu )}{\xi -n-\ldots} }} =\xi -\re^{\alpha (1+\nu )} \left\{\sum _{n=0}^{\infty }\frac{[\alpha (1+\nu )]^{n} }{n!\, (\xi +\alpha -n)}  \right\}^{-1} .
\end{equation}
The same calculation of ${\Large\textsl{m}}_{1}^{+} $ (with all lines
\includegraphics[width=0.65in]{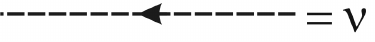}) yields two equal expressions
\begin{equation} \label{GrindEQ__14_}
{\Large\textsl{m}}_{1}^{+} (\xi )=\frac{\alpha \nu }{\xi+1-\displaystyle\frac{2 \alpha \nu }{\xi +2-\ldots-\displaystyle\frac{n \alpha \nu }{\xi +n-\ldots} } } =\xi -\re^{\alpha \nu } \left[\sum _{n=0}^{\infty }\frac{(\alpha \nu )^{m} }{n!\, (\xi +\alpha +n)}  \right]^{-1} .
\end{equation}
We should note that the representation of ${\Large\textsl{m}}_{1}^{\pm } (\xi )$ in the form of a chain fraction converges to the exact value at all $\xi $ and $\alpha $ faster than that in the form of the sum over $n$. We should also note that as it is shown in \cite{Tka84} and proven by formulae  \eqref{GrindEQ__11_}--\eqref{GrindEQ__14_} at $T=0$~K when $\nu=0$ MO $\Large\textsl{m}(\xi,\nu=0)=\Large\textsl{m}_{1}^{-} (\xi,\nu=0)$ has the exact representation in the form of a continuous chain fraction, the $n$-th and $(n+1)$-th links of which are related by the expression $\Large\textsl{m}_{n}(\xi)=\alpha(n+1)[\xi-(n+1)-\Large\textsl{m}_{n+1}(\xi)]^{-1}$. It is interesting that the qualitatively similar continuous fractions appeared in the other non-perturbative theories for quasi-particles interacting with phonons, in particular, in the momentum average approximation for inhomogeneous systems, \cite{Ebr12}.

The terms of MO \eqref{GrindEQ__11_}, which describe the mixed sequences of scattering processes
\begin{equation} \label{GrindEQ__15_}
{\Large\textsl{m}}_{2l} (\xi )={\Large\textsl{m}}_{2l}^{-}(\xi )+{\Large\textsl{m}}_{2l}^{+}(\xi ),\qquad {\Large\textsl{m}}_{2l+1} (\xi )= {\Large\textsl{m}}_{2l+1}^{-} (\xi )+{\Large\textsl{m}}_{2l+1}^{+}(\xi ),
\end{equation}
have the following analytical form:
\begin{equation} \label{GrindEQ__16_}
{\Large\textsl{m}}_{2l}^{\pm} (\xi )=[\alpha^{2} \nu (1+\nu )]^{l} \sum _{n=0}^{\infty }\alpha ^{2n}  \left\{\begin{array}{c} {1+\nu } \\ {\nu } \end{array}\right\}^{2n} f_{2(n+l)}^{(l+1)\mp } (\xi ),   \qquad  (l=1,2,\ldots),
\end{equation}
\begin{equation} \label{GrindEQ__17_}
{\Large\textsl{m}}_{2l+1}^{\mp } (\xi )=\alpha ^{2l+1} \left\{\begin{array}{c} {\nu } \\ {1+\nu } \end{array}\right\}^{l} \left\{\begin{array}{c} {1+\nu } \\ {\nu } \end{array}\right\}^{l+1} \sum _{n=0}^{\infty }\alpha ^{2n}  \left\{\begin{array}{c} {1+\nu } \\ {\nu } \end{array}\right\}^{2n} f_{2(n+l+{1 \mathord{\left/{\vphantom{1 2}}\right.\kern-\nulldelimiterspace} 2} )}^{(l+1)\mp } (\xi ),\qquad (l=1,2,\ldots).
\end{equation}

Here, the functions $f(\xi)$ are consistently defined by the diagrams with differently directed phonon lines of the order 2$l$  and 2$l$+1, respectively. Thus, for example, the mixed diagrams of the second order ($l$=1), define MO ${\Large\textsl{m}}_{2}^{2\mp}(\xi )$  and the respective functions  $f_{2}^{2(\mp )} (\xi )$
\begin{equation} \label{GrindEQ__18_}
\raisebox{0pt}{\hfil\includegraphics[width=4.4in]{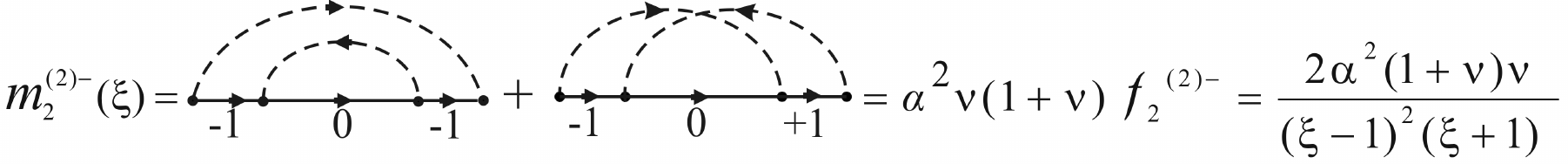}}\raisebox{10pt}{$\,\,,$}
\end{equation}
\begin{equation} \label{GrindEQ__19_}
\raisebox{0pt}{\hfil\includegraphics[width=4.4in]{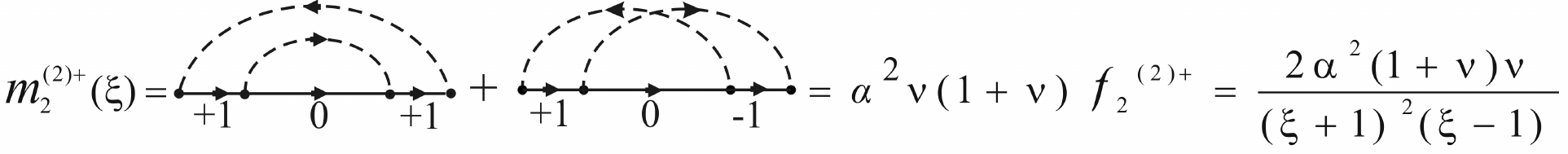}}\raisebox{10pt}{$\,\,.$}
\end{equation}
Each MO of the third order [${\Large\textsl{m}}_{3}^{(3)\mp}(\xi )$] contains 30 diagrams with three differently directed phonon lines. Thus, to avoid cumbersomeness, we present the final analytical expressions
\begin{equation} \label{GrindEQ__20_}
{\Large\textsl{m}}_{3}^{(3)\mp } (\xi )=\alpha ^{3} \left\{\begin{array}{c} {\nu } \\ {1+\nu } \end{array}\right\}\left\{\begin{array}{c} {1+\nu } \\ {\nu } \end{array}\right\}^{2} f_{3}^{(3)\mp } (\xi ),
\end{equation}
where
\begin{equation} \label{GrindEQ__21_}
f_{3}^{(3)\mp } (\xi )=\frac{4}{(\xi \mp 1)\, ^{2} (\xi \pm 1)\, } \left[\frac{2}{(\xi \mp 1)\, (\xi \mp 2)\, } +\frac{1}{(\xi \pm 1)^{2} \, } \right]+\frac{2}{(\xi \mp 1)^{3} \, } \left[\frac{1}{(\xi \mp 2)\, \, } +\frac{2}{(\xi \pm 1)} \right]^{2}.
\end{equation}
The exact analytical expressions for the functions  $f(\xi )$ of the higher order are clearly determined by diagram technique, although the number of diagrams rapidly increases for the bigger order.

In general case, when there is no decay, the renormalized energy spectrum of the system of quasi-particles interacting with phonons is obtained  from the solutions of dispersional equation
\begin{equation} \label{GrindEQ__22_}
\xi -{\Large\textsl{m}}(\xi )=0.
\end{equation}

In order to study the properties of such systems in the model (a) with the condition ($\widehat{n}^{2} =\widehat{n}$) and in general model (b), we are going to use  ${\Large\textsl{m}}_{\text a} (\xi )$ and ${\Large\textsl{m}}_{\text{b}} (\xi )$ MO, respectively. Herein, for the model (a), we use the known \cite{Dav76} analytical expression for the Green's function and Dyson equation in dimensionless variables and obtain the analytical expression for  ${\Large\textsl{m}}_{\text{a}} (\xi )$
\begin{equation} \label{GrindEQ__23_}
{\Large\textsl{m}}_{\text{a}} (\xi )=\xi -\re^{\alpha (1+2\nu )} \left[\sum _{n,\, p=0}^{\infty }\frac{\alpha ^{n+p} (1+\nu )^{n} \nu ^{p} }{n!p!\, (\xi +\alpha -n+p+\ri\eta )}  \right]^{-1},
\end{equation}
which is correct at arbitrary values of  $\alpha $ and $\nu $.

In the general model (b) we use the MO
\begin{equation} \label{GrindEQ__24_}
{\Large\textsl{m}}_{\text{b}} (\xi )={\Large\textsl{m}}_{1} (\xi )+{\Large\textsl{m}}_{\text m} (\xi ), \qquad {\Large\textsl{m}}_{\text m} (\xi )={\Large\textsl{m}}_{2}^{(2)} (\xi )+{\Large\textsl{m}}_{3}^{(3)} (\xi ),
\end{equation}
where ${\Large\textsl{m}}_{1} (\xi )$ is the just defined MO which describes the unmixed sequences of the scattering processes which is correct for the arbitrary values $\alpha $, $\nu $ and $\xi $ as well as ${\Large\textsl{m}}_{\text m} (\xi )$ which describes the mixed sequences of the scattering processes of the second and third order. This is sufficient for the majority of bulk crystals and nano-structures where the magnitude of the electron-phonon interaction energy is not bigger than the phonon energy and the temperature is in the range of the room one, so that $\alpha ,\nu <1$.

Besides the renormalized energy spectrum of the system in both models, we are going to analyze the so-called phonon ``coats'' of quasi-particles, namely the average number of phonons  ($N_{\text{a,\,b}} $) surrounding the quasi-particles in the respective states depending on the values $\alpha $ and $\nu $. It is known \cite{Lev73} that the average numbers of phonons ($N_{n} $) surrounding the quasi-particles in the states with the energies  $\xi _{n} $ obtained from the dispersional equation \eqref{GrindEQ__22_} in general case are defined as
\begin{equation} \label{GrindEQ__25_}
N_{n} =\left[\frac{\partial {\Large\textsl{m}}(\xi)}{\partial \xi } \Bigg|_{\xi =\xi _{n} }-1\right]^{-1}.
\end{equation}
As far as the spectrum of renormalized energies $E_{n}$ is exactly fixed by formula \eqref{GrindEQ__2_} in the model (a), $N_{\text{a}n} $ is also calculated analytically exactly
\begin{equation} \label{GrindEQ__26_}
N_{\text{a}\{ n\geqslant 0\} } =1-\re^{-\alpha (1+2\nu )} \frac{\alpha ^{\left|n\right|} }{\left|n\right|!} \left\{\begin{array}{c} {1+\nu } \\ {\nu } \end{array}\right\}^{\left|n\right|},\qquad(n=0,\pm 1,\pm 2,\ldots).
\end{equation}
In the model (b), the complete MO contains the term  ${\Large\textsl{m}}_{1} (\xi )$, which in analytical form describes all unmixed scattering processes independently of the values $\alpha $ and $\nu $ and of the rest of terms, which describe the mixed scattering processes, where besides the MOs defined in the second and third order, the higher orders demand the accounting of the bigger number of diagrams. Thus, in the model (b), we are going to analyze $N_{\text{b}} $ for the system with weak coupling ($\alpha <1$)
\begin{equation} \label{GrindEQ__27_}
N_{\text{b}} (\xi _{\text{b}n} )=\left\{\frac{\partial \left[{\Large\textsl{m}}_{1} (\xi)+{\Large\textsl{m}}_{\text m} (\xi)\right]}{\partial \xi } \Bigg|_{\xi =\xi _{\text{b}n} } -1\right\}^{-1},
\end{equation}
where $\xi _{\text{b}n} $ are the solutions of dispersional equation \eqref{GrindEQ__22_} with MO \eqref{GrindEQ__24_}.

As far as both MOs ${\Large\textsl{m}}_{1} (\xi )$ and ${\Large\textsl{m}}_{\text{a}} (\xi )$ are right at arbitrary $\alpha $ and $\nu $, in order to compare with $N_{\text{a}}$, we are going to study the properties of $N_{1}(\xi_{\gtrless0})$ numbers, which can be analytically written as
\vspace{-1.5mm}
\begin{equation} \label{GrindEQ__28_}
N_{1} (\xi _{n_{\gtrless}0} )=1+\left(1-\frac{\re^{\alpha (1+\nu )} \displaystyle\sum _{p=0}^{\infty }\frac{[\alpha (1+\nu )]^{p} }{p![\xi _{n} +\alpha (1+\nu )-p]^{2} }  }{\left\{\displaystyle\sum _{p=0}^{\infty }\frac{[\alpha (1+\nu )]^{p} }{p![\xi _{n} +\alpha (1+\nu )-p]}  \right\}^{2} } +\frac{\re^{\alpha \nu } \displaystyle\sum _{p=0}^{\infty }\frac{(\alpha \nu )^{p} }{p!(\xi _{n} +\alpha \nu +p)^{2} }  }{\left[\displaystyle\sum _{p=0}^{\infty }\frac{(\alpha \nu )^{p} }{p!(\xi _{n} +\alpha \nu +p)}  \right]^{2} } \right)^{-1} ,
\end{equation}
where  $\xi _{n} $ are the solutions of the equation \eqref{GrindEQ__22_} with MO ${\Large\textsl{m}}_{1} (\xi )$.
\pagebreak

The developed theory gives the opportunity to calculate, reveal and analyze the common and different features of the renormalized spectra of the systems of quasi-particles interacting with polarization phonons in both models.

\section{Analysis of the formation of the renormalized spectra of the system of localized quasi-particles interacting with polarization phonons}
 The temperature evolution and the reasons of differences between renormalized spectra of the system of localized quasi-particles interacting with polarization phonons in two models are advisable to be studied comparing the respective mass operators as functions of dimensionless energy $\xi $ varying in the range $-3\leqslant \xi \leqslant 3$ at different $\alpha $ and $\nu$.

\begin{figure}[!t]
\begin{minipage}[h]{.49\linewidth}
\centering{\includegraphics[width=\linewidth]{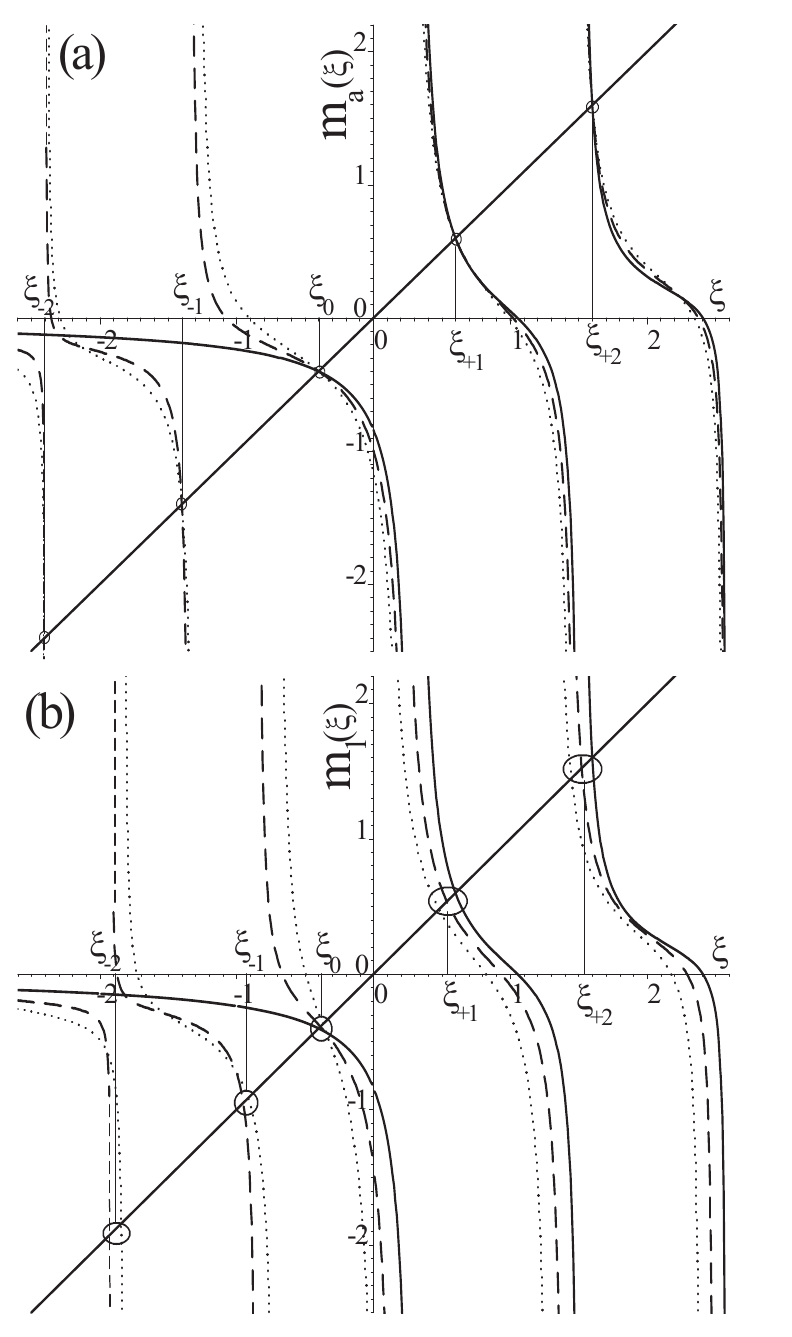}}
\caption{MO ${\Large\textsl{m}}_{\text{a}}$ (a) and ${\Large\textsl{m}}_{1}$ (b) as functions of $\xi$ and of the  values of energies $\xi_{n}$ at $\nu$=0 (---), 0.2~(-~-~-), 0.4 ($\cdot\cdot\cdot$) and  $\alpha=0.4$.} \label{fig-mal2}
\end{minipage}
\begin{minipage}[h]{.49\linewidth}
\vspace{-2mm}
\centering{\includegraphics[width=\linewidth]{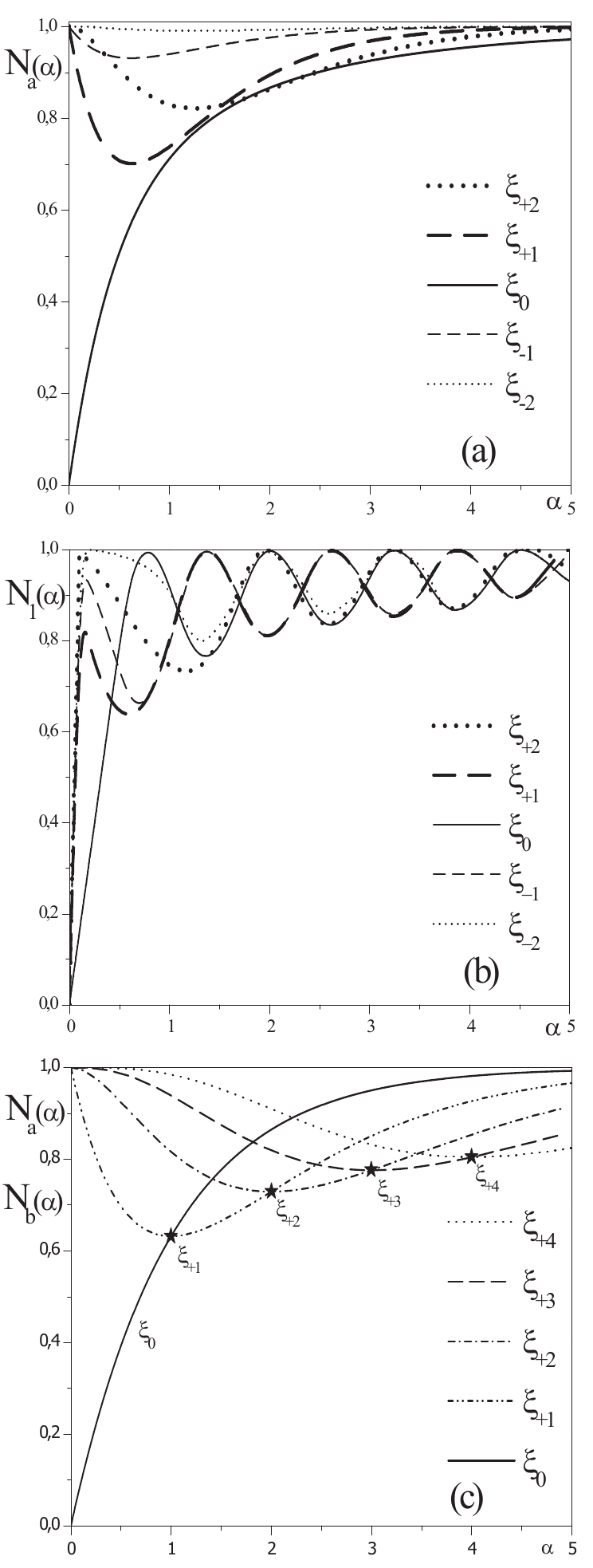}}
\vspace{-2mm}
\caption{The average number of phonons $N_{\text a}$ (a) and $N_{1}$ (b) as functions of the coupling constant $\alpha$ at $\nu=0.3$ and $N_{\text{a}}=N_{\text{b}}=N_{1}$ at $\nu=0$ (c).
} \label{fig-mal1}
\end{minipage}
\vspace{-2mm}
\end{figure}

The typical dependences of MO $\Large\textsl{m}_{\text{a}} $, model (a), and MO terms $\Large\textsl{m}_{1} $ of unmixed processes, model (b), are shown in figure~\ref{fig-mal2} as functions of $\xi $. Both of them are defined by the exact analytical expressions which are correct at arbitrary $\alpha $ and  $\nu $. Figure~\ref{fig-mal2} proves that both functions are qualitatively similar and at $T=0$~K when $\nu =0$, are even completely equal \cite{Dav76, Tka03, Tka84}. However, at $T\neq0$~K when $\nu \ne 0$, their behavior is different in actual regions of energy $\xi $ where the solutions of dispersion equations [$\xi _{n(\text{a,\,b})} $] define the energy spectra. When $\nu =0$ , from \eqref{GrindEQ__11_}, \eqref{GrindEQ__15_}--\eqref{GrindEQ__17_} it is clear that MO $\Large\textsl{m}_{\text m} \equiv 0$, thus $\Large\textsl{m}_{\text{a}} (\xi )\left|_{\nu =0} \right. =\Large\textsl{m}_{\text{b}} (\xi )\left|_{\nu =0} \right. =\Large\textsl{m}_{1} (\xi )\left|_{\nu =0} \right.$, since the renormalized energy spectra in both models are the same ($\xi _{n} =-\alpha +n$, $n=0,1,2,\ldots$) and equidistant. They are formed by the interaction between quasi-particles and virtual polarization phonons in the processes of their gradual creation with further gradual annihilation.

From figure~\ref{fig-mal2} it is clear that at the increasing temperature ($\nu $), model (b), in the approximated MO $\Large\textsl{m}_{1} (\xi )$ where only unmixed sequences of scattering processes are taken into account, all modes of $\Large\textsl{m}_{1} (\xi )$, which form the bound states of quasi-particles with one, two and more phonons with the energies \linebreak ($\xi _{+1},\xi _{+2},\ldots$), shift into the low-energy region while the mode which forms the ground renormalized state~($\xi _{0} $) and all states without one, two and more phonons ($\xi _{-1} ,\xi _{-2},\ldots$) in the vicinity of the energies of these states --- vice versa. Consequently, as it is clear from figure~\ref{fig-mal2}~(b) and table~\ref{tbl1}, the renormalized spectrum of the system becomes not equidistant if the temperature increases. In this approximation, in the model (b), the same states as in the model (a) are observed. However, the properties of temperature dependence of the spectra and, as it is revealed, the properties of average phonon numbers ($N$) in phonon ``coats'' of quasi-particles are different.

\begin{table}[!h]
\caption{The values of energies $\xi_{n}$ at  $\nu$=0, 0.2, 0.4 and $\alpha$=0.4.} \label{tbl1} %\vspace{2ex}
\begin{center}
\renewcommand{\arraystretch}{0}
\begin{tabular}{|c||c|c|c|c|c|}
\hline\hline \raisebox{-0.4ex}[0pt]{$\nu$}&
\raisebox{-0.4ex}[0pt]{$\xi_{-2}$}& \raisebox{-0.4ex}[0pt]{$\xi_{-1}$}& \raisebox{-0.4ex}[0pt]{$\xi_{0}$}& \raisebox{-0.4ex}[0pt]{$\xi_{+1}$}& \raisebox{-0.4ex}[0pt]{$\xi_{+2}$}\strut\\
\hline
\rule{0pt}{2pt}&&&&&\\
\hline
\raisebox{-0.3ex}[0pt]{0}&\raisebox{-0.3ex}[0pt] {$-$}& \raisebox{-0.3ex}[0pt]{$-$}& \raisebox{-0.3ex}[0pt]{$-0.4$}& \raisebox{-0.3ex}[0pt]{0.6}& \raisebox{-0.3ex}[0pt]{1.6}\strut\\
\hline
\raisebox{-0.3ex}[0pt]{0.2}&\raisebox{-0.3ex}[0pt] {$-1.92$}& \raisebox{-0.3ex}[0pt]{$-0.95$}& \raisebox{-0.3ex}[0pt]{$-0.38$}& \raisebox{-0.3ex}[0pt]{0.537}& \raisebox{-0.3ex}[0pt]{1.522}\strut\\
\hline
\raisebox{-0.3ex}[0pt]{0.4}&\raisebox{-0.3ex}[0pt]{$-1.84$}& \raisebox{-0.3ex}[0pt] {$-0.92$}& \raisebox{-0.3ex}[0pt]{$-0.36$}& \raisebox{-0.3ex}[0pt]{0.480}& \raisebox{-0.3ex}[0pt]{1.446}\strut\\
\hline\hline
\end{tabular}
\renewcommand{\arraystretch}{1}
\end{center}
\end{table}

The average number of phonons in phonon ``coats'' for different states of quasi-particles with the energies $\xi _{n} $ are presented in figure~\ref{fig-mal1} as functions of the coupling constant $\alpha $ at $\nu $=0, 0.3. If $\nu $=0, these numbers~($N$) are the same in both models and are obtained  exactly analytically
\begin{equation} \label{GrindEQ__29_}
N_{\text{a}} (\xi _{n} ,\alpha )=N_{\text{b}} (\xi _{n} ,\alpha )=N_{1} (\xi _{n} ,\alpha )=1-\re^{-\alpha } \frac{\alpha ^{n} }{n!}\,,\qquad  (n=0,1,2,\ldots).
\end{equation}
They have an interesting property: in each $n$-th state of the system, these numbers as functions of $\alpha $ have one minimum at $\alpha =n$, where, figure~\ref{fig-mal1}~(c), their magnitudes (marked by stars) are equal to the magnitudes of the numbers of the neighbour $(n+1)$ state
\begin{equation} \label{GrindEQ__30_}
N(\xi _{n} ,\alpha =n)=N_{\text{b}} (\xi _{n+1} ,\alpha =n)=1-\frac{n^{n} }{n!} \re^{-n},\quad (n=1,2,3,\dots ).
\end{equation}
The average phonon numbers in each $n$-th state of quasi-particle asymptotically increase till one in the region $\alpha \geqslant 1$. From figure~\ref{fig-mal1}~(c) it is clear that at $\alpha \leqslant 1$ the number of phonons in the ``coat'' decreases only for the renormalized ground state of quasi-particle when $\alpha $ becomes smaller.

At non-zero temperatures ($\nu\neq0$), besides the high-frequency states with the energies $\xi _{n\geqslant {\rm 1}} $, one can observe the low-frequency ones with the energies $\xi _{n\leqslant-1} $ in both models. Now, the average phonon numbers ($N_{\text{a}}, \, N_{1} $) as functions of $\alpha $ are essentially different in two models. These numbers were calculated using formulae \eqref{GrindEQ__26_}--\eqref{GrindEQ__28_}, and the results are shown in figures~\ref{fig-mal1}~(a), (b), respectively.  From figure~\ref{fig-mal1}~(a) it is clear that in the model (a) the numbers $N_{\text{a}} $ in quasi-particles phonon ``coat'' depend on $\alpha $  in the similar way as it was for $\nu $=0, but the positions of their minima are now fixed by the relationship
\begin{equation} \label{GrindEQ__31_}
\alpha _{n} =\frac{\left|n\right|}{1+2\nu }\,,\qquad(n=\pm1,\pm2,\ldots ),
\end{equation}
and the magnitudes  $\min N_{\text{a}} (\pm n)$  are fixed by the expression
\begin{equation} \label{GrindEQ__32_}
\min N_{\text{a}} (n)=1-5^{-\left|n\right|} \frac{\left|n\right|^{\left|n\right|} }{\left|n\right|!} \frac{1}{(1+2\nu )^{\left|n\right|} } \left(\begin{array}{c} {1+\nu } \\ {\nu } \end{array}\right)^{\left|n\right|},\qquad(n=\pm1,\pm2,\ldots ).
\end{equation}

Thus, when the temperature increases, the magnitude $\alpha _{\gtrless0} $ also decreases. Herein,  $\min N_{n\leqslant -1} $ decreases as well, while $\min N_{n\geqslant 1} $ increases. In all low-frequency states at all $\alpha $, figure~\ref{fig-mal1}~(a), $N_{n\leqslant -1} $ are nearly low maximal ($N_{n\leqslant -1} \approx 1$) but $N_{n\geqslant 0} $  weakly shift to the direction of smaller $\alpha $ with respect to their analogues at $\nu $=0.

The average numbers of phonons in the ``coats'' of quasi-particles are presented in figure~\ref{fig-mal1}~(b) as functions of $\alpha $ for several states of the system in the model (b). This figure proves that in the region $\alpha\geqslant1$ the numbers $N_{1}$ $(n=0,\pm 2,\pm 4,\ldots)$ oscillate in  antiphase with the numbers $N_{1}$ $(n=\pm 1,\pm 3,\ldots)$. These oscillations are quasi-periodical and, according to the calculations, their period and amplitude decrease when $\nu $ increases. Comparing figure~\ref{fig-mal1}~(a) with figure~\ref{fig-mal1}~(b), one can see that in the approximation where in the complete MO only all unmixed processes [$\Large\textsl{m}_{1} (\xi )$] are accounted at $\alpha \geqslant1$, the average numbers of phonons in quasi-particles ``coats'' [$N_{1} (n)$] in model (b) correlate well with their analogues [$N_{\text{a}} (n)$] in model (a).

\begin{figure}[!t]
\centerline{\includegraphics[width=1\textwidth]{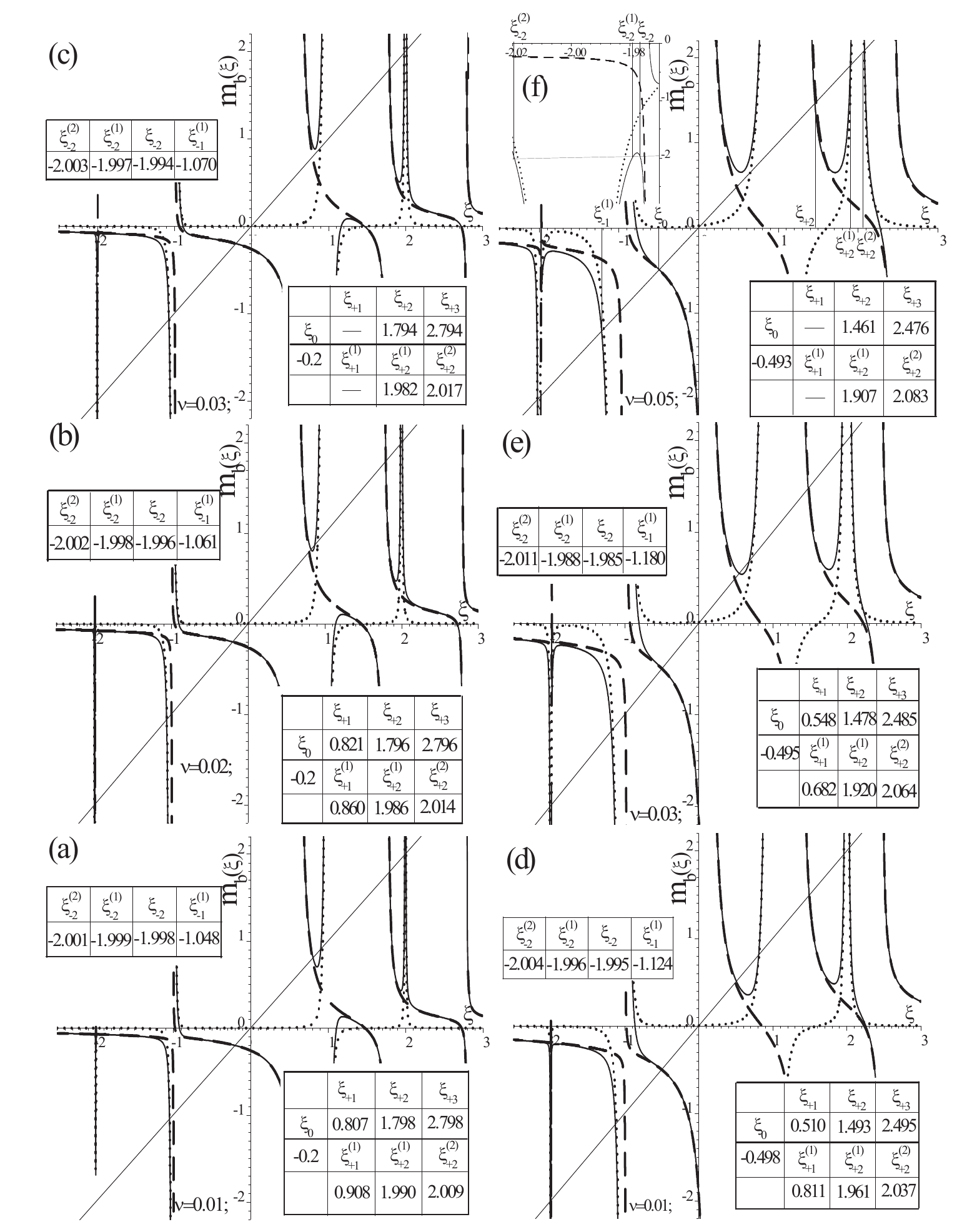}}
\caption{MO ${\Large\textsl{m}}_{\text{b}}(\xi)$ (---) and its terms ${\Large\textsl{m}}_{1}$ (- - -), ${\Large\textsl{m}}_{\text m}$ ($\cdot\cdot\cdot$) as functions of $\xi$ at $\alpha=0.2$~(a,b,c), $\alpha=$0.5~(d,e,f). The values of energies are presented in the insert.}
\label{fig3}
\end{figure}

Taking into account the analysis of the properties of the spectra and average numbers of phonons in the ``coats'' of localized quasi-particles interacting with polarization phonons we can conclude that the results of the exact model (a) according to their main features are the same as the results of model (b) in which all unmixed scattering processes are considered.

Now we are going to study the properties of the spectra and average phonon numbers in the states of the system of localized quasi-particles interacting with phonons in model (b), where in the complete MO $\Large\textsl{m}_{\text{b}} (\xi )$, besides the unmixed processes described by MO $\Large\textsl{m}_{1}(\xi) $, the mixed scattering processes of the second and third orders over the coupling constant  are considered.

The dependences of MO $\Large\textsl{m}_{\text m}(\xi)$ at different $\alpha$ and $\nu$ are presented in figure~\ref{fig3}, which shows the formation and behaviour of the energy spectrum of the system in general model (b). Figures~\ref{fig3}~(a)--(f) prove that besides the states revealed in the model (a) and in the model (b) without taking into account the mixed processes (the modified or even the disappeared ``old'' states), one can see completely ``new'' states, their energies being also obtained from the solution of dispersional equation \eqref{GrindEQ__22_}. Of course, the separation of the states at ``old'' and at ``new'' bound states of the system is fully conditional, because all states equally belong to the system. However, such separation makes it possible to better reveal the reasons of a different behaviour of these states. In figure~\ref{fig3} one can identify the origin of all states of the system fixed by the solutions of equation \eqref{GrindEQ__22_}.

The analysis of the terms $\Large\textsl{m}_{1}^{\pm } (\xi )$ and $\Large\textsl{m}_{\text m}^{\pm } (\xi )$ of the complete MO at $\nu\neq0$ shows, figure~\ref{fig3}, that renormalized energy spectrum of the system, as a set of solutions of dispersional equation, is mainly formed by those of them that contain the poles in the respective region of energies $\xi $. The analytical form of MO \eqref{GrindEQ__11_} proves that both its terms $\Large\textsl{m}_{1} (\xi )$ and $\Large\textsl{m}_{\text m} (\xi )$ are essential only in the vicinity of $\xi\sim\xi _{0} \,$, while in high-frequency region, the prevailing contribution into the formation of the spectrum is performed by the terms $\Large\textsl{m}_{1}^{-} (\xi )$ and $\Large\textsl{m}_{\text m}^{-} (\xi )$ and in low-frequency region --- $\Large\textsl{m}_{1}^{(+)} (\xi )$ and $\Large\textsl{m}_{\text m}^{(+)} (\xi )$, respectively.  In the structures of MO $\Large\textsl{m}_{1}^{-} $ and $\Large\textsl{m}_{\text m}^{-} $ there prevail the terms proportional to  (1+$\nu $), describing the processes accompanied by creation of phonons, which are exactly the ones that form the high-frequency bound states of the system. In the structures of MO $\Large\textsl{m}_{1}^{(+)} (\xi )$ and $\Large\textsl{m}_{1}^{-} (\xi )$ there prevail the terms proportional to  $\nu $, describing the processes accompanied by annihilation of phonons, which are exactly the ones that form the low-frequency bound states of the system.

Figure~\ref{fig3} also shows that in the regions of energies $\xi $, where in the vicinity of the solutions of dispersional equation, the condition  $\left|\Large\textsl{m}_{1} (\xi )\right|>\left|\Large\textsl{m}_{\text m} (\xi )\right|$ is fulfilled, the ``old'' states are formed and at the condition $\left|\Large\textsl{m}_{\text m} (\xi )\right|>\left|\Large\textsl{m}_{1} (\xi )\right|$ --- the ``new'' states appear. From the physical considerations and according to the previous analysis one can conclude that the ``old'' states with renormalized energies $\xi _{n} =0,\pm 1,\pm 2,\ldots$ are mainly formed by unmixed scattering processes and the ``new'' ones with renormalized energies $\xi _{+1}^{(1)}$, $\xi _{+2}^{(2)}$ are formed by the mixed processes [figure~\ref{fig3}~(f)]. When the solutions of dispersional equation $\Large\textsl{m}_{\text m}(\xi )=\xi $ exist and the condition $\Large\textsl{m}_{1}(\xi )$ = $\Large\textsl{m}_{\text m}(\xi )=\xi /2$ is fulfilled, there appear ``new'' degenerated states of the system with renormalized energies $\overline{\xi _{n} }$ and with equal contributions of mixed and unmixed scattering processes.

\begin{figure}[!b]
\vspace{-3mm}
\centerline{\includegraphics[width=0.82\textwidth]{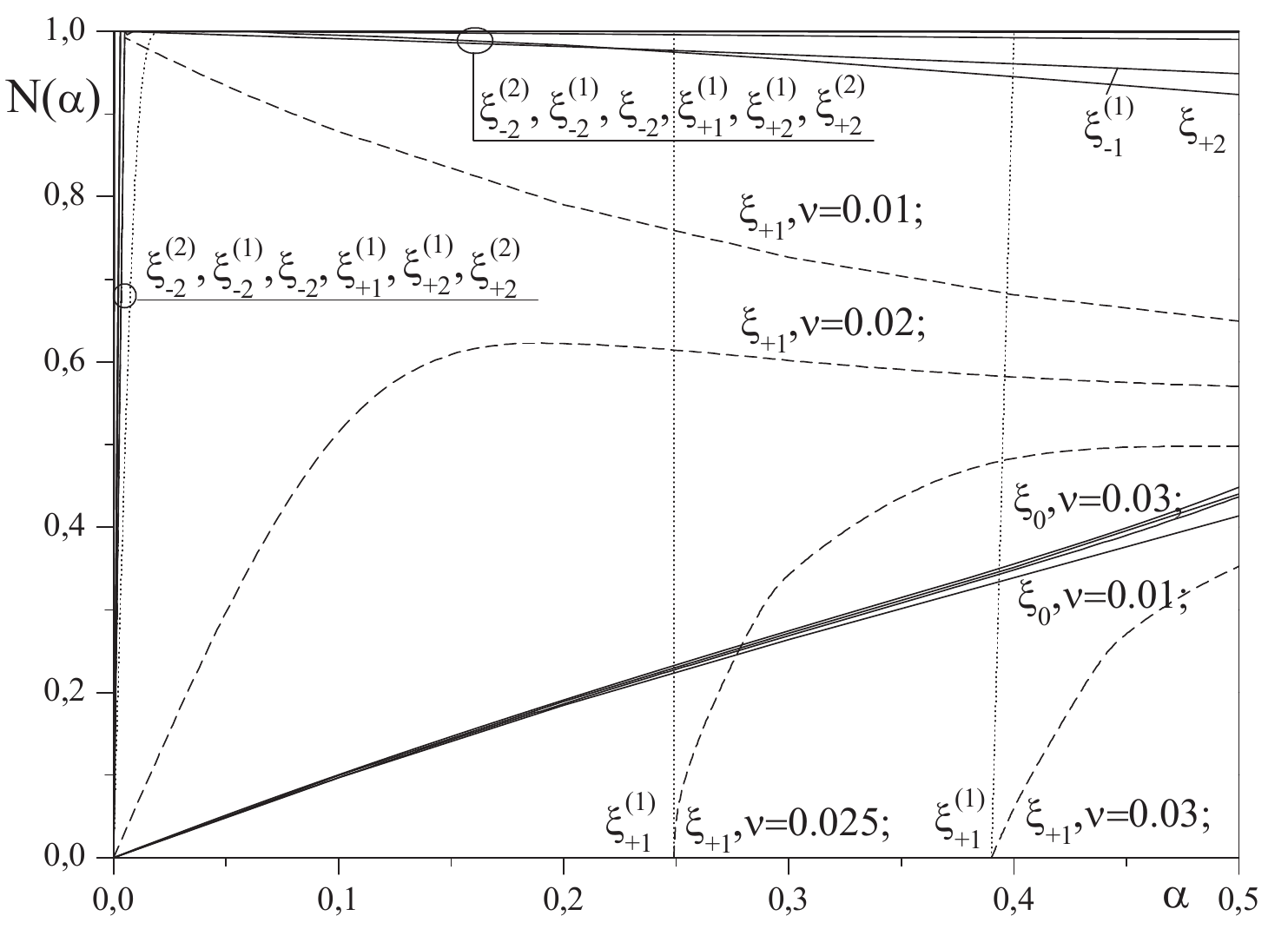}}
\caption{Average phonon numbers ($N$) in phonon ``coats'' of quasi-particles as functions of the coupling constant $\alpha$ in several bound states at different temperatures ($\nu$).}
\label{fig4}
\end{figure}

Moreover, from figure~\ref{fig3} one can clearly see the properties of renormalized energies of quasi-particles interacting with polarizational phonons (in the range $-3\leqslant \xi \leqslant 3$ ) depending on the coupling constant ($0\leqslant \alpha \leqslant 0.5$) and temperatures being in the vicinity of room ones ($0<\nu <0.1$). The energy of the ground state of the system  $\xi _{0} \approx -\alpha $ is shifted into the low-frequency region. Together with the energies ($\xi _{+1} ,\xi _{+2} ,\xi _{+3} $) of ``old'' high-frequency bound states it forms a quasi-equidistant spectrum with the difference between the energies being of one phonon order which weakly depends on the values $\alpha $ and $\nu $. Up to three ``new'' bound states can appear in the high-frequency region. One of them has the energy $\xi _{+1}^{(1)} $ ($\xi _{+1} \leqslant \xi _{+1}^{(1)} <\xi _{+2}^{(2)} $) and the other two with the energies $\xi _{+2}^{(1)} $ and $\xi _{+2}^{(1)} $ ($\xi _{+2} \leqslant \xi _{+2}^{(1)}$; $\xi _{+2}^{(2)} <\xi _{+3} $). Depending on the relationships between the values $\alpha $ and $\nu $, all high-frequency states can exist, degenerate and disappear. Figure~\ref{fig3} proves that at the fixed $\alpha $, the existing high-frequency states vary in such a way that the differences between the energies of neighboring ``new'' and ``old'' states ($\xi _{+1}^{(1)} -\xi _{+1} $) or ($\xi _{+2}^{(1)} -\xi _{+2} $) decrease when the temperature ($\nu $) increases. At certain temperature, $\overline{T}(\overline{\nu})$, the difference between the energies of one-phonon states $\xi _{+1}^{(1)} =\xi _{+1} $ disappears since they degenerate into one state, which  further disappears at a higher temperature. When temperature still increases, the pair of neighboring states with the energies $\xi _{+2}^{(1)} $ and $\xi _{+2} $ show a similar behaviour.

Comparing figure~\ref{fig3}~(a) and \ref{fig3}~(b) one can see that the degeneration of the respective pairs of the energy levels with a further disappearance of the respective states is observed at a stronger interaction between quasi-particles and phonons ($\alpha $) and at higher temperature ($\nu $). This clear from physical considerations because the stronger is quasi-particles-phonons coupling, the harder it is to break it, thus, the temperature should increase.

Contrary to the high-frequency bound states, the low-frequency ones with the energies $\xi _{-1}^{(1)} $, $\xi _{-2} $, $\xi _{-2}^{(1)} $, $\xi _{-2}^{(2)} <\xi _{0} $ exist only at $T\neq0$~K. Their temperature dependence differs as well, figure~\ref{fig3}. One can see that in the vicinity of the energy $\xi\sim1$, only one ``new'' state with the energy $\xi _{-1}^{(1)} $ is formed while the ``old'' one with the energy $\xi _{-1} $ is absent because the contribution of unmixed scattering processes into  MO prevails. In the vicinity  $\xi\sim2$, three bound states are observed: one ``old'' ($\xi _{-2} $) and two ``new'' ($\xi _{-2}^{(1)} $, $\xi _{-2}^{(2)} $), presented at the upper insert in figure~\ref{fig3}. This figure and the tables prove that the low-frequency spectrum of energies weakly changes at increasing temperature independently of $\alpha $.

The revealed differences in the behaviour of energy spectra of bound states of the system of quasi-particles interacting with phonons in two models are also seen in the dependences of average phonon numbers ($N$) in the phonon ``coats'' on $\alpha $ and $\nu $. Figure~\ref{fig4} shows that at the regime of weak coupling ($0\leqslant\alpha\leqslant0.5$) at the fixed $\alpha $, the increasing temperature ($\nu $) strongly decreases the number $N$ only for the high-frequency state with the energy $\xi _{+1} $. This fact correlates with the obtained temperature dependences of neighbouring energy levels  $\xi _{+1} $ and $\xi _{+1}^{(1)} $ , starting from their co-existence till degeneration and disappearance. The average number of phonons in the ground state $\xi _{0} $ weakly decreases at an increasing temperature while in all other states, the numbers $N$ almost do not depend on $\nu $. Herein, in the vicinity of very small $\alpha $, they sharply increase when the interaction increases.

\section{Main results and conclusions}

Using the method of Feynman-Pines diagram technique, we developed the theory of renormalized spectra of localized quasi-particles interacting with polarization phonons at finite temperatures in a wide range of energies containing the energies of the bound states of the system. The properties of renormalized spectrum of the system are analyzed and compared in two models: in the known model \cite{Dav76} in which the problem is solved exactly when the additional condition fulfills  ($\widehat{n}^{2}=\widehat{n}$ , where $\widehat{n}$ is the operator of quasi-particles number) and in the general model with the regime of a weak coupling between quasi-particles and phonons.

It is established that, contrary to the model with an additional condition, where there always exist bound stationary states of the system with equidistant energy spectrum independent of temperature, in the general model, the renormalized spectrum is found to be much richer and more complicated. This is caused by different formation of bound states in mixed and in unmixed sequences of the processes of quasi-particle scattering while interacting with phonons.

It is shown that in the general model, the unmixed processes of quasi-particle scattering, accompanied by the creation of phonons, form a group of bound states with quasi-equidistant spectrum while the mixed processes form a group of states with non-equidistant spectrum in high-frequency region. The states located near the ground state are mainly formed by the mixed processes of quasi-particle scattering with annihilation of phonons causing a non-equidistant spectrum in the low-energy region.

It is revealed that in the systems with weak coupling, if the temperature increases to the room one, the location of the energy levels weakly changes while in high-frequency region they gradually degenerate in pairs and then disappear.

Finally, we should note that the use of the computer method of construction and calculation of the higher order diagrams, which describe the mixed sequences of the scattering processes of quasi-particles interacting with phonons, would essentially broaden the range of energies containing the bound states of the system and outrun the frames of the weak coupling. Besides, the further development of Feynman-Pines diagram technique would help to solve the well-known urgent problems regarding the existence and properties of the excited states of electron-phonon systems (in particular, polaron) at arbitrary temperatures.

%
%% If you have problems with typesetting in ukrainian uncomment lines below.
%
% \lastpage
% \end{document}

\ukrainianpart

\title{Властивості і температурна еволюція спектру системи локалізованих квазічастинок взаємодіючих з поляризаційними фононами у двох моделях}
\author{М.В. Ткач, Ю.О. Сеті, О.М. Войцехівська, О.Ю. Питюк}
\address{Чернівецький національний університет ім. Ю.~Федьковича,\\ вул. Коцюбинського, 2, 58012 Чернівці, Україна}

\makeukrtitle

\begin{abstract}
\tolerance=3000%
Методом діаграмної техніки Фейнмана-Пайнса розраховано і досліджено енергетичний спектр локалізованих квазічастинок, взаємодіючих з поляризаційними фононами  у широкій області енергій при скінченній температурі системи.
Показано, що загальна модель системи крім зв'язаних станів, відомих зі спрощеної моделі з накладеною додатковою умовою на оператор повного числа квазічастинок, навіть зі слабким зв'язком містить нові зв'язані стани. Виявлена роль багатофононних процесів у формуванні перенормованого спектру системи.  Встановлено причини появи, еволюції та зникнення окремих пар зв'язаних станів у залежності від величини енергії зв'язку та температури.
\keywords квазічастинка, фонон, функція Гріна, масовий оператор

\end{abstract}

\end{document}